\documentclass[pdflatex,sn-nature]{sn-jnl}

\usepackage[T1]{fontenc}
\usepackage{graphicx}%
\usepackage{multirow}%
\usepackage{amsmath,amssymb,amsfonts}%
\usepackage{amsthm}%
\usepackage[title]{appendix}%
\usepackage{xcolor}%
\usepackage{textcomp}%
\usepackage{manyfoot}%
\usepackage{booktabs}%
\usepackage{algorithm}%
\usepackage{algorithmicx}%
\usepackage{algpseudocode}%
\usepackage{listings}%
\usepackage{makecell}
\usepackage{booktabs}
\usepackage{placeins}
\usepackage{needspace}
\usepackage{siunitx}

\hypersetup{
  pdftitle={Scensory: Real-Time Robotic Olfactory Perception for Joint Identification and Source Localization},
  pdfauthor={Yanbaihui Liu; Erica Babusci; Claudia K. Gunsch; Boyuan Chen},
  pdfsubject={Robotic olfaction and biological source localization},
  pdfkeywords={robotic olfaction, electronic nose, volatile organic compounds, source localization}
}

\theoremstyle{thmstyleone}%
\theoremstyle{thmstyletwo}%

\theoremstyle{thmstylethree}%

\newcommand{\projectwebsite}{%
  \begin{center}
    \small\url{http://generalroboticslab.com/Scensory}
  \end{center}
}

\makeatletter
\patchcmd{\@maketitle}{\printabstract}{\projectwebsite\printabstract}{}{\errmessage{patch failed}}
\makeatother

\begin{document}
\title{Scensory: Real-Time Robotic Olfactory Perception for Joint Identification and Source Localization}


\author*[1]{\fnm{Yanbaihui} \sur{Liu}}\email{yanbaihui.liu@duke.edu}

\author[2]{\fnm{Erica} \sur{Babusci}}\email{erica.babusci@duke.edu}

\author[2]{\fnm{Claudia K.} \sur{Gunsch}}\email{ckgunsch@duke.edu}


\author*[1,3,4]{\fnm{Boyuan} \sur{Chen}}\email{boyuan.chen@duke.edu}

\affil[1]{\orgdiv{Department of Mechanical Engineering and Materials Science}, \orgname{Duke
University}, \orgaddress{\city{Durham}, \state{NC}, \country{USA}}}

\affil[2]{\orgdiv{Department of Civil and Environmental Engineering}, \orgname{Duke
University}, \orgaddress{\city{Durham}, \state{NC}, \country{USA}}}

\affil[3]{\orgdiv{Department of Electrical and Computer Engineering},\orgname{Duke
University}, \orgaddress{\city{Durham}, \state{NC}, \country{USA}}}

\affil[4]{\orgdiv{Department of Computer Science}, \orgname{Duke
University}, \orgaddress{\city{Durham}, \state{NC}, \country{USA}}}


\abstract{
Olfaction offers robots access to chemical information that is largely inaccessible to vision, touch, and audition, yet using airborne chemical signals for spatial perception remains challenging because local volatile organic compound (VOC) measurements are shaped by complex chemical transport and sensor dynamics. We introduce \textit{Scensory}, a robotic olfaction framework that learns to jointly infer biological source identity and relative location from short temporal VOC measurements. Using a robot-automated data collection platform, we pair VOC dynamics from cross-sensitive gas sensor arrays with spatial supervision and train models to predict fungal identity, source direction, and distance. We show that a single sensor array can extract all three quantities from only 3\,s of local measurements under ambient environmental conditions, achieving species classification accuracy of up to 80.13\%, directional accuracy of up to 68.65\%, and mean absolute distance errors of 0.110--0.131\,m. Incorporating measurements from multiple spatial locations further reduces ambiguity, improving peak species and directional accuracies by 9.72 and 18.66 percentage points, respectively. We then embody this learned olfactory perception on a mobile robot, where successive local predictions acquired during motion are transformed into a world-frame evidence map, allowing observations from different positions and headings to reinforce persistent source hypotheses and guide closed-loop localization. Across eight selected indoor runs, the robot achieves a planar endpoint error of $0.606\pm0.294$\,m. Our results establish airborne chemical dynamics as a viable perceptual signal for robots to recognize biological sources, reason about their spatial origin, and autonomously navigate toward them under ambient environments.
}




\maketitle

\section{Introduction}\label{sec1}

Autonomous robots use sensory observations to perceive their surroundings and guide action. Vision supports object perception, geometric reconstruction, and navigation \cite{redmon2016you, campos2021orbslam3, localization2023}. Acoustic sensing can reveal spatial relationships from naturally occurring or interaction-generated sounds \cite{liu2024sonicsense, liu2025wildfusion, liu2026sonicfly}. Tactile sensing provides complementary information about contact geometry and force \cite{yuan2017gelsight, do2023densetact}. These modalities often couple observations closely to environmental geometry or physical interactions, providing spatial information that robots can use directly for action. Robotic olfaction, however, presents a fundamentally different perception problem. Airborne chemical signals, especially in ambient environments, are transported through diffusion and airflow, producing measurements that are intermittent, temporally delayed, and only indirectly related to the location of their source \cite{celani2014odor,zhang2026advanced}. Therefore, a robot may detect a chemical signature without directly knowing where it originated. These challenges are particularly consequential for biological sources such as indoor fungal growth, where complex mixtures of volatile organic compounds (VOC) need to support not only recognition, but also spatial reasoning and action.

Fungal contamination provides an important setting in which such chemical perception could enable autonomous inspection, continuous monitoring, and early warning. Exposure to dampness and indoor mold is associated with adverse respiratory and allergic health outcomes \cite{mendell2011respiratory, tischer2011association, caillaud2018indoor}, but existing detection workflows are poorly suited for continuous, spatially resolved deployment. Culture-dependent assays require manual sample collection, incubation, and laboratory identification, while microscopy and molecular methods still generally depend on collected samples and offline processing \cite{pitkaranta2011molecular, guarner2011histopathologic, halliday2015molecular}. Turnaround can range from days for culture-based analysis to tens of hours for molecular assays. For example, an environmental air-sampling study of \textit{Aspergillus fumigatus} reported approximately seven days for culture-based results and 48 hours from collection to qPCR results \cite{bellanger2010detection}. Moreover, identifying the physical origin of contamination introduces an additional challenge, often requiring building-scale investigation, assessment of moisture damage, and targeted examination or sampling of suspect materials \cite{epaMoldInvestigation}. A robotic system capable of sensing fungal emissions directly while moving through an environment can instead couple detection with spatial search.

VOC sensing provides a promising basis for such a capability. Fungi emit diverse volatile compounds during metabolism, including alcohols, ketones, and sulfur-based compounds, that can form species- and lifestyle-associated chemical patterns \cite{buzzini2005production, aisala2019odor, sommer2021wild, guo2021volatile}. Electronic nose (eNose) systems \cite{persaud1982analysis} use cross-sensitive sensor arrays and statistical or machine learning models to distinguish these patterns and have been studied extensively in food, agricultural, and environmental applications \cite{li2024electronic, cabanes2009early, liu2018discrimination, suchorab2019method}. Under controlled conditions, eNose systems have achieved high classification performance for selected fungal-infection and processing conditions \cite{makarichian2024use} and distinguished fungal-infected samples from healthy controls across different contamination levels \cite{Borowik2024}. However, some eNose workflows for fungal detection use enclosed headspace accumulation before measurement \cite{tan2023volatile,baietto2010evaluation}, while sorbent-based pre-concentration has also been used for mycotoxin-related screening \cite{piergiovanni2025electronic}, making them hard to deploy directly in real-world conditions. Moreover, these studies mainly focus on \textit{what} is present, but they not generally provide the relative direction and distance needed for a robot to determine \textit{where} the emitting source is located.

In parallel, robotic odor source localization has investigated how mobile agents can infer and navigate toward the spatial origin of airborne chemicals using plume tracking, concentration gradients, probabilistic search, airflow information, and learning-based policies \cite{hayes2002distributed, loutfi2009gas, marjovi2014multi, hassan2024robotic}. More recent systems have introduced bio-inspired and multimodal strategies that adapt robot behavior to olfactory observations or combine chemical sensing with complementary perception modalities \cite{monroy2018semantic, zhang2025oevs, fukui2025biohybrid, shigaki2026insect}. These methods demonstrate that chemical sensing can support robotic navigation when spatial information can be recovered from plume structure, concentration variation, airflow, or additional sensing. Biological VOC sources in ambient indoor environments pose a different challenge. Local measurements may be chemically informative while providing weak and ambiguous spatial cues without strong concentration gradients or structured airflow as in prior assumptions. It becomes much more challenging when a robot needs to reason about both source identity and location from those observations. Hence, such a capability has remained missing.

Here, we introduce \textit{Scensory}, a learning-based robotic olfaction framework that addresses the joint problem of identifying a biological source and estimating its location from short, local chemical observations in ambient indoor environments. Scensory couples chemical sensing, robot-generated spatial self-supervision, and sequential embodied perception. It learns from short temporal VOC measurements paired automatically with source identity, relative direction, and distance, allowing a single sensor array to infer semantic and spatial properties from a 3\,s chemical observation. We first characterize how much source information is recoverable from a single local array and then use synchronized measurements from six spatially distributed arrays to examine how additional spatial context reduces chemical ambiguity. We subsequently embody the learned local perception on a mobile robot, where successive observations acquired from different positions and headings are transformed into a common world-frame evidence map. Instead of acting on any individual prediction, our framework enables the robot to accumulate mutually consistent evidence to generate and verify source hypotheses, navigate toward candidate locations, and perform local chemical confirmation. Our experiments establish a perception--action framework in which robot motion transforms weak and ambiguous local VOC measurements into persistent spatial evidence, enabling biological source recognition and autonomous localization from airborne chemical dynamics.


\section{Results}\label{sec2}

\subsection{Robot-automated data collection grounds fungal VOC measurements in space}

\begin{figure}[ht]
    \centering
    \includegraphics[width=1\linewidth]{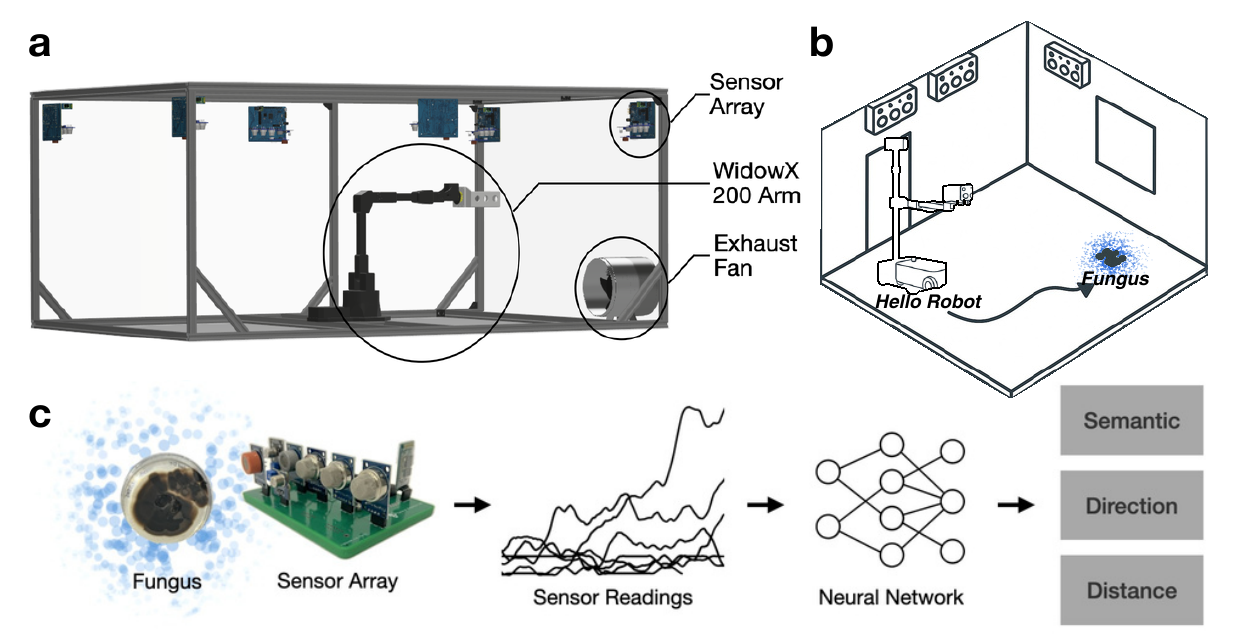}
    \caption{$|$ \textbf{Scensory overview and VOC-based fungal inference workflow.} 
    \textbf{a}, Robot-automated data collection platform (1.5\,m × 1.0\,m × 0.6\,m) with six spatially distributed VOC sensor arrays and a robotic arm that positions fungal sources throughout the workspace. An exhaust fan clears residual volatiles between trials.
    \textbf{b}, Scensory sensing configurations. The mobile robot carries a single array for source localization. Fixed arrays illustrate a possible distributed configuration used here to evaluate the value of additional spatial context.
    \textbf{c}, Inference configuration. Short temporal VOC measurements are used to estimate fungal identity, relative source direction, and source distance. Blue shading schematically represents VOC dispersal.}

    \label{fig:system_overview}
\end{figure}

Scensory combines cross-sensitive VOC sensing with learning-based inference of biological source identity and relative location. The sensing unit integrates six metal-oxide semiconductor (MOx) gas sensors, an electrochemical formaldehyde sensor, and temperature and humidity sensors into a compact array (Fig.~\ref{fig:system_overview};Sec.~\ref{sec:method-sensor-array}). Because individual gas sensors respond to overlapping sets of compounds but with different sensitivities, source information is encoded in the joint temporal response across channels instead of any single sensor \cite{persaud1982analysis}. A custom printed circuit board (PCB) integrates the sensing elements, onboard microcontroller, and wireless communication hardware (Sec.~\ref{sec:method-sensor-array}). Sensor measurements are transmitted via Bluetooth to a Raspberry Pi 5 running ROS for synchronized data acquisition, model inference, and robot coordination. During inference, Scensory operates on 3\,s of measurement windows to estimate fungal identity, relative source direction, and source distance, enabling continuous local chemical perception from short temporal observations.

Learning spatially grounded chemical perception requires training data in which VOC measurements are paired with accurate source identity and relative source position. To generate such data systematically, we developed a robot-automated data collection platform (Fig.~\ref{fig:system_overview}, Sec.~\ref{sec:method-platform}). A WidowX 200 robotic arm carries a petri dish containing a fungal culture sample through a predefined sequence of spatial waypoints inside a custom acrylic testbed. At each waypoint, the source remains stationary for 10--20\,s while six fixed Scensory arrays record synchronously. The robot end-effector pose provides spatial ground truth, allowing each VOC observation to be associated with the corresponding source location. An exhaust fan is activated between trials to purge residual VOCs, and sampling resumes after sensor readings return to baseline.

The synchronized recordings support two complementary sensing conditions. In the primary single-array setting, each array is treated independently and source coordinates are expressed in its local reference frame, matching the information available to a mobile robot carrying one Scensory array (Fig.~\ref{fig:system_overview}c). The same recordings also permit a spatial-context analysis in which measurements from all six locations are processed jointly. This second condition provides a reference for determining how simultaneous spatial observations reduce the ambiguity of local chemical measurements.

\begin{figure}[ht]
    \centering
    \includegraphics[width=0.8\textwidth]{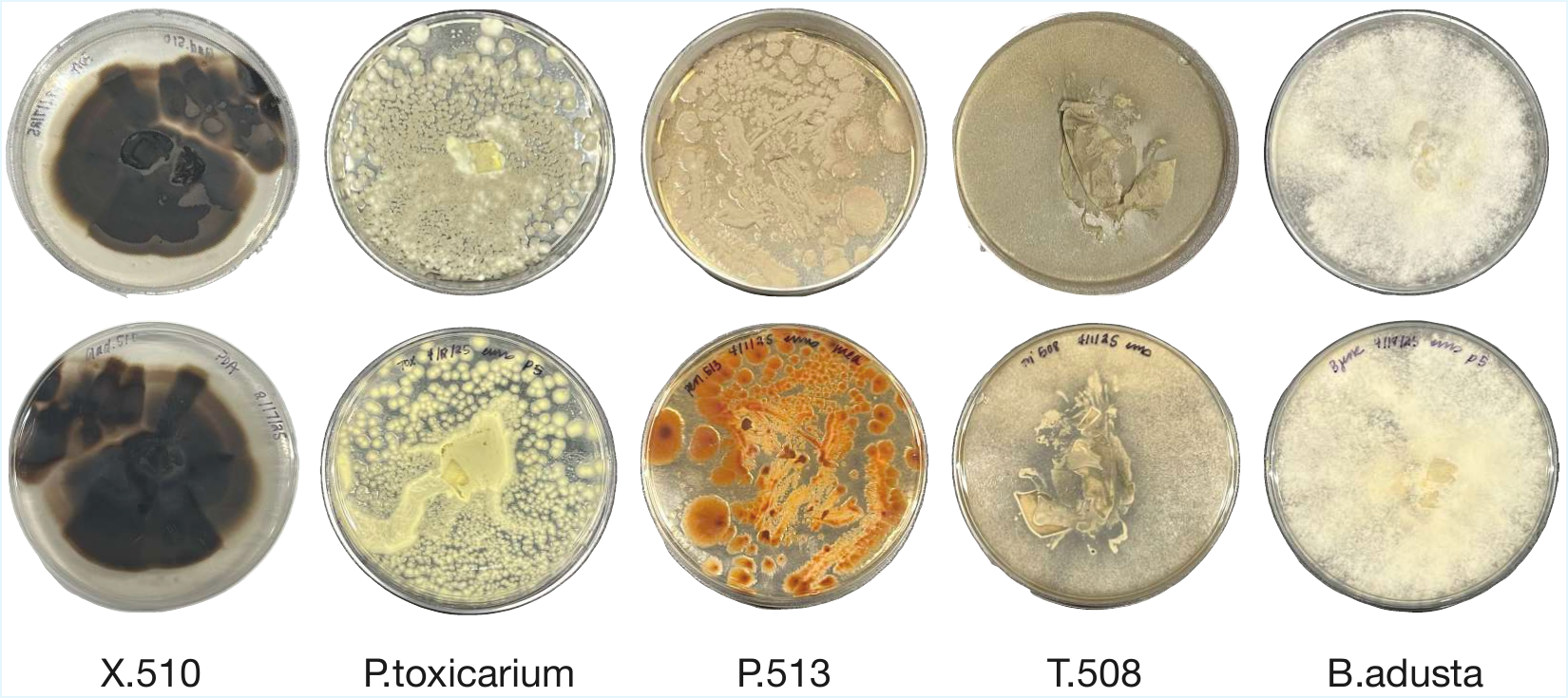}
    \caption{$|$ \textbf{Fungal isolates used for Scensory evaluation.}
    Top and bottom views of representative colonies cultured on 90\,mm plates. From left to right: \textit{Xylaria cubensis.510} (\textit{X.510}), \textit{Penicillium toxicarium} (\textit{P. toxicarium}), \textit{Penicillium.513} (\textit{P.513}), \textit{Trichoderma.508} (\textit{T.508}), and \textit{Bjerkandera adusta} (\textit{B. adusta}).
    }
    \label{fig:fungi_panel}
\end{figure}

We evaluated Scensory using five fungal isolates: \textit{Bjerkandera adusta} (\textit{B. adusta}), \textit{Penicillium toxicarium} (\textit{P. toxicarium}), \textit{Xylaria cubensis.510} (\textit{X.510}), \textit{Penicillium.513} (\textit{P.513}), and \textit{Trichoderma.508} (\textit{T.508}) (Fig.~\ref{fig:fungi_panel}). The isolates span multiple genera and environmental origins, providing chemically diverse biological sources for evaluating both semantic discrimination and spatial inference.

\subsection{Local VOC observations encode source identity and relative location}

We first asked how much semantic and spatial information can be recovered from a single local chemical observation. Each prediction was restricted to measurements from one Scensory array, with species, direction, and distance inferred from a 3\,s VOC sequence. Models were evaluated across five independent training runs to characterize variability in learned performance.

\begin{figure}[htbp]
    \centering
    \includegraphics[width=1\linewidth]{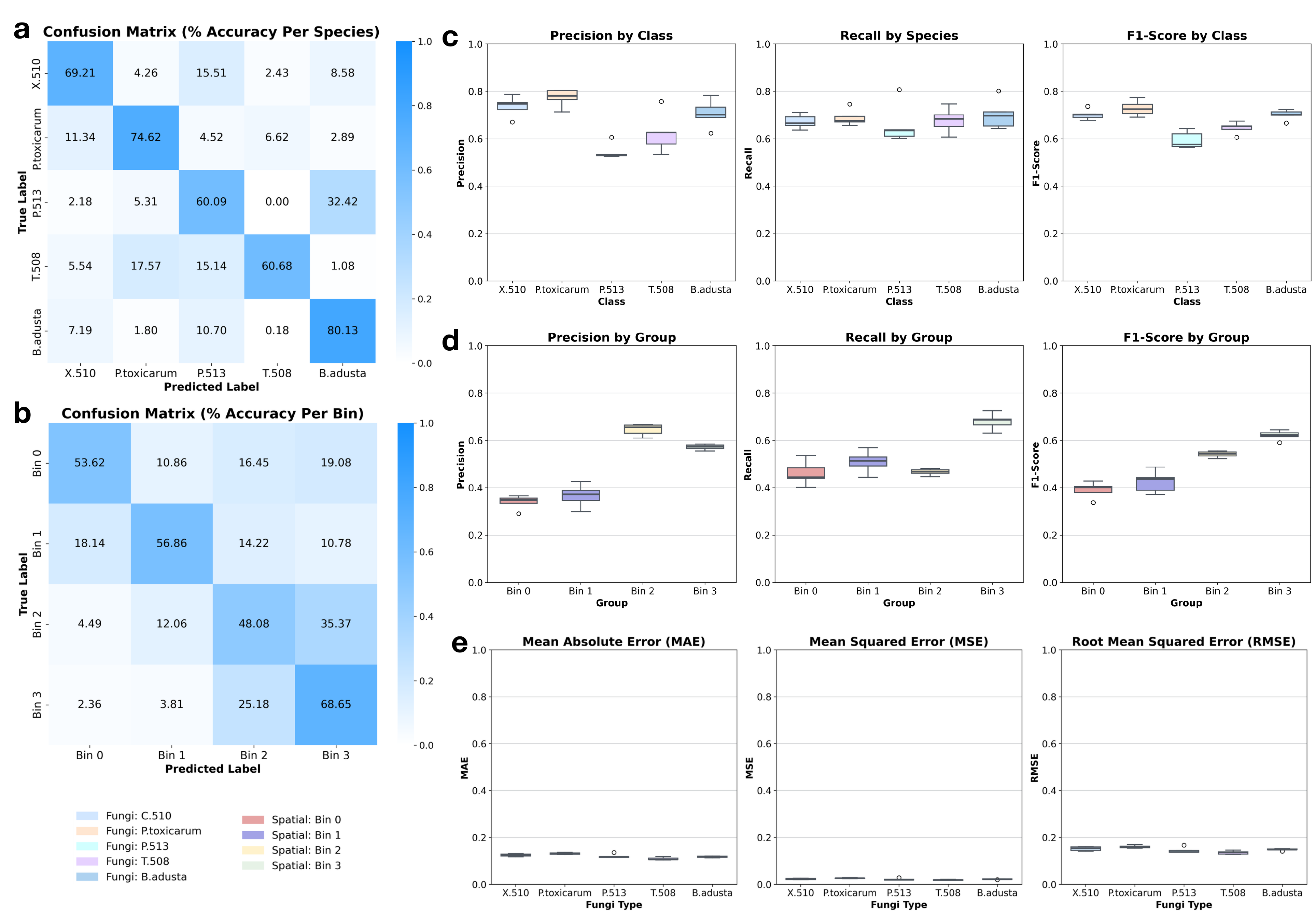}
    \caption{$|$ \textbf{Single-array inference of fungal identity and relative source location.}
    \textbf{a}, Confusion matrix for fungal species classification. 
    \textbf{b}, Confusion matrix for source-direction classification across four array-relative directional bins.
    \textbf{c}, Precision, recall, and F1-score for species classification across five fungal types. 
    \textbf{d}, Precision, recall, and F1-score for source-direction classification.
    \textbf{e}, Distance-regression errors (MAE, RMSE, MSE) for each fungal species.}
    \label{fig:singlearray_performance}
\end{figure}

\subsubsection{Fungal species identification}

Single-array species classification achieved class-wise accuracies ranging from 60.09\% to 80.13\% across the five fungal isolates (Fig.~\ref{fig:singlearray_performance}a). Thus, a 3\,s VOC observation from one sensing location retained substantial information about source identity despite the absence of spatially distributed measurements. The confusion matrix revealed several structured error patterns. \textit{X.510} was frequently confused with \textit{P.513}, \textit{P.513} showed confusion with \textit{B. adusta}, and \textit{T.508} was occasionally classified as \textit{P.513} or \textit{P. toxicarium}. These errors suggest that local chemical observations are informative but not uniquely species-specific.

These confusions are consistent with the fact that fungal species can share VOCs while differing in other components of their volatile profiles \cite{guo2021volatile}, providing a plausible biological basis for overlapping sensor-level signatures. At the same time, cross-sensitive sensor arrays distinguish odorants through combined response patterns rather than requiring individual channels to identify specific chemicals \cite{persaud1982analysis}. Scensory therefore distinguishes fungal sources through distributed response patterns across the sensor array, while partially overlapping VOC mixtures can produce similar local representations.

Performance was reproducible across independent training runs, although discrimination difficulty remained species dependent (Fig.~\ref{fig:singlearray_performance}c). Precision and recall generally ranged from approximately 0.55 to 0.80, further indicating that local VOC observations support species identification, although some species remained harder to distinguish than others.

\subsubsection{Source direction and distance estimation}

We next tested whether the same local VOC observations contain information about where the source is located relative to the sensor. Direction classification was more challenging than species identification, with per-bin accuracy ranging from 48.08\% to 68.65\% across the four directional bins (Fig.~\ref{fig:singlearray_performance}b), compared with a 25\% chance level. Unlike direct geometric sensing, a single chemical array does not observe source bearing explicitly. Directional information must instead be inferred from temporal patterns produced by chemical transport and the cross-sensitive sensing response. Chemical transport can produce intermittent local odor observations \cite{celani2014odor}. Transport variability and overlapping sensor responses may therefore contribute to this ambiguity. Nevertheless, the above-chance performance shows that short local VOC dynamics contain recoverable information about source direction despite substantial ambiguity.

Performance varied among directional bins (Fig.~\ref{fig:singlearray_performance}d). Bin~3 showed the strongest overall performance with comparatively balanced precision and recall, whereas Bin~2 showed higher precision and lower recall, consistent with more conservative predictions. Such asymmetries are consistent with the spatial and temporal variability expected from local airborne chemical transport \cite{celani2014odor}.

Local VOC dynamics also provided information about source range. Each regressor was evaluated on test samples from its corresponding ground-truth species, so these errors quantify distance estimation conditional on species identity. Species-specific distance regressors achieved mean absolute errors between 0.1097\,m and 0.1311\,m across the five fungal isolates (Fig.~\ref{fig:singlearray_performance}e; Table~\ref{tab:distance_errors}). The lowest error was obtained for \textit{T.508} (0.1097 $\pm$ 0.0060\,m), whereas \textit{P. toxicarium} showed the highest error (0.1311 $\pm$ 0.0041\,m). To compare the distance errors with the physical source size, we additionally defined a source-size-normalized resolution ratio using the radius of the fungal culture plate. Fungal samples were cultivated in 90\,mm-diameter Petri dishes, corresponding to a plate radius of 45\,mm. We define

\[
\mathrm{Resolution\ Ratio}
=
\frac{r_{\mathrm{plate}}}{\mathrm{MAE}},
\]

where $r_{\mathrm{plate}}=0.045$\,m. A ratio below one indicates that the average distance-estimation error exceeds the plate radius, whereas larger values indicate error scales that are smaller relative to the physical source extent.

Across species, the resulting resolution ratios ranged from 0.34 to 0.41 (Table~\ref{tab:distance_errors}), indicating that the average distance-estimation error exceeds the plate radius but remains on the same order of magnitude as the physical source size. The error scale is comparable to dimensions reported for robotic cleaning systems, such as a 0.2,m spray standoff or a 30\,cm$\times$30\,cm mopping pad \cite{lee2020automated,pookkuttath2023snail}. Additional observations acquired during closed-loop motion can subsequently refine this approximate range cue, as examined in our mobile experiments below. These above results establish that a single 3\,s chemical observation provides a coarse but measurable estimate of source range in addition to source identity and direction.

\begin{table}[ht]
\centering
\caption{\textbf{Distance regression errors across fungal isolates.} Mean absolute error (MAE) and root mean squared error (RMSE) are reported in meters, and mean squared error (MSE) is reported in square meters. Values report the mean and standard deviation across independent training runs. Resolution Ratio is defined as the Petri-dish radius (0.045\,m) divided by MAE, providing a source-size-normalized measure of distance-estimation error.}
\label{tab:distance_errors}
\begin{footnotesize}
\begin{tabular}{l@{\hskip 8pt}c@{\hskip 4pt}c@{\hskip 8pt}c@{\hskip 4pt}c@{\hskip 8pt}c@{\hskip 4pt}c@{\hskip 8pt}c}
\toprule
\textbf{Species} & \makecell{MAE\\(mean) [m]} & \makecell{MAE\\(std) [m]} & \makecell{MSE\\(mean) [m$^2$]} & \makecell{MSE\\(std) [m$^2$]} & \makecell{RMSE\\(mean) [m]} & \makecell{RMSE\\(std) [m]} & \makecell{Resolution \\ Ratio} \\
\midrule
\textit{X.510}          & 0.1246 & 0.0050 & 0.0239 & 0.0021 & 0.1513 & 0.0079 & 0.36 \\
\textit{P. toxicarium}  & 0.1311 & 0.0041 & 0.0258 & 0.0016 & 0.1601 & 0.0058 & 0.34 \\
\textit{P.513}          & 0.1198 & 0.0092 & 0.0198 & 0.0038 & 0.1388 & 0.0119 & 0.38 \\
\textit{T.508}          & 0.1097 & 0.0060 & 0.0182 & 0.0018 & 0.1358 & 0.0073 & 0.41 \\
\textit{B. adusta}      & 0.1169 & 0.0033 & 0.0217 & 0.0015 & 0.1467 & 0.0036 & 0.39 \\
\bottomrule
\end{tabular}
\end{footnotesize}
\end{table}

\subsubsection{Comparison with baseline methods}

We compared Scensory with three alternative inference methods using the same 3\,s input windows and identical train, validation, and test data as shown in Table~\ref{tab:baseline}. The baselines included a Drosophila-inspired denoise--aggregate--MLP pipeline~\cite{yue2024drosophila}, a static-feature MLP using per-channel mean, standard deviation, and maximum values, and a Random Forest trained on the same summary statistics. These comparisons test whether explicitly learning temporal VOC structure provides advantages over representations that collapse each observation into static statistics.

For species identification, Scensory achieved the highest overall performance among the evaluated methods, with 69.94\% accuracy and a 68.97\% F1-score. The static MLP was the strongest baseline, reaching 66.23\% accuracy and a 65.24\% F1-score, followed by Random Forest and the Drosophila-inspired pipeline. These results suggest that modeling temporal VOC dynamics provides a measurable advantage over static summary features for fungal species discrimination. Directional prediction was more challenging. Random Forest achieved the highest accuracy at 58.31\%, slightly exceeding Scensory at 57.00\%. However, Scensory attained higher recall and F1-score, reaching 56.75\% and 52.48\%, respectively. This trade-off indicates that Scensory produces more balanced predictions across directional bins, even though its overall accuracy is marginally lower. These results suggest that the task-specific temporal encoder with attention captures discriminative structure beyond what can be recovered from summary statistics or denoise-aggregate pipelines.

\begin{table}[ht]
\centering
\caption{Single-array baseline comparison for species classification and spatial bin prediction.}
\label{tab:baseline}
\scriptsize
\setlength{\tabcolsep}{4pt}
\renewcommand{\arraystretch}{1.0}
\begin{tabular}{
l l
S[table-format=2.2]
S[table-format=2.2]
S[table-format=2.2]
S[table-format=2.2]
}
\toprule
\textbf{Method} & \textbf{Task} & {\textbf{Acc(\%)}} & {\textbf{P(\%)}} & {\textbf{R(\%)}} & {\textbf{F1(\%)}} \\
\midrule
\multirow{2}{*}{Drosophila}
  & Species & 36.67 & 38.65 & 40.32 & 37.75 \\
  & Directions    & 33.61 & 31.25 & 35.80 & 30.35 \\
\addlinespace[1pt]
\multirow{2}{*}{Static MLP}
  & Species & 66.23 & 66.70 & 66.28 & 65.24 \\
  & Directions    & 45.06 & 38.24 & 45.72 & 38.88 \\
\addlinespace[1pt]
\multirow{2}{*}{Random Forest}
  & Species & 59.83 & 61.62 & 61.03 & 58.73 \\
  & Directions    & {\bfseries 58.31} & {\bfseries 51.39} & 52.36 & 50.33 \\
\addlinespace[1pt]
\multirow{2}{*}{\textbf{Scensory (ours)}}
  & Species & {\bfseries 69.94} & {\bfseries 70.01} & {\bfseries 68.95} & {\bfseries 68.97} \\
  & Directions    & 57.00 & 50.49 & {\bfseries 56.75} & {\bfseries 52.48} \\
\bottomrule
\end{tabular}%
\end{table}

\subsection{Spatially distributed measurements reduce chemical ambiguity}

The single-array experiments revealed that local VOC observations contain both semantic and spatial information, but also substantial ambiguity. We therefore asked how inference changes when the same chemical field is observed simultaneously from multiple locations. Synchronized measurements from all six Scensory arrays were processed jointly as a spatial-context reference.

\begin{figure}[htbp]
    \centering
    \includegraphics[width=0.95\linewidth]{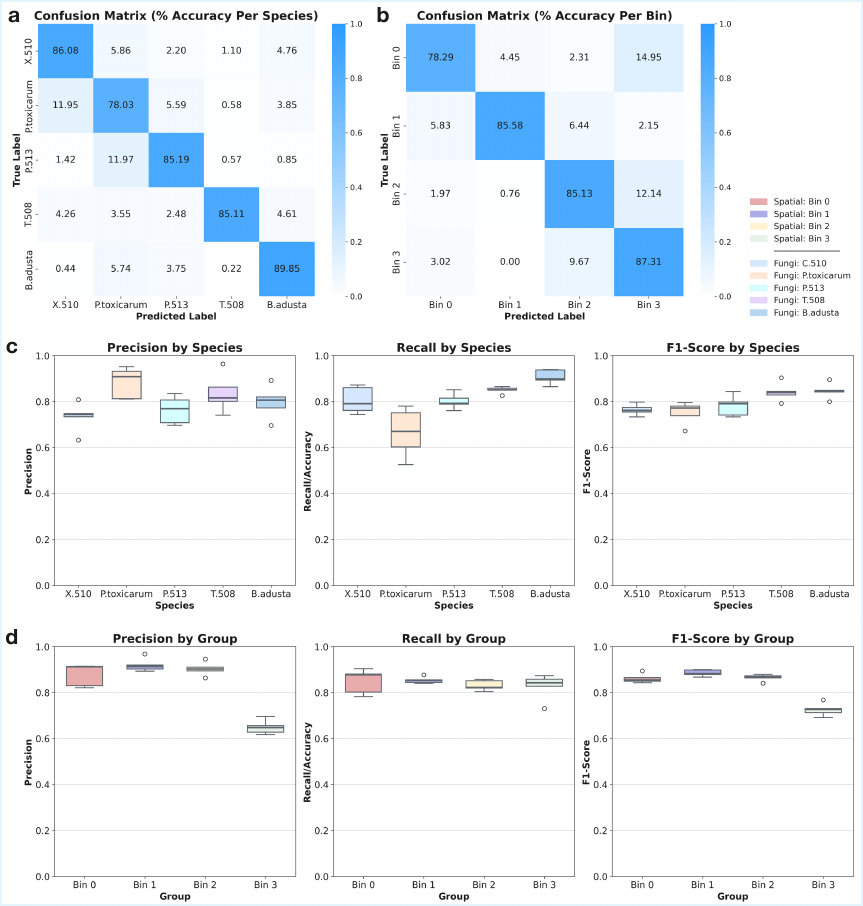}
    \caption{$|$ \textbf{Spatially distributed VOC measurements improve semantic and directional inference.}
    \textbf{a}, Confusion matrix for species classification. Rows indicate ground-truth species and columns indicate model predictions. Values represent row-normalized classification accuracy. 
    \textbf{b}, Confusion matrix for spatial localization. Rows indicate true directional quadrants and columns represent predicted quadrants. 
    \textbf{c}, Precision, recall, F1-score for species classification across five independent trainings. Box plots indicate median and interquartile ranges; whiskers represent score range, and outliers are shown as circles. 
    \textbf{d}, Performance metrics for spatial localization across four directional bins.}
    \label{fig:multiarray_performance}
\end{figure}

For fungal identification, spatial context increased class-wise accuracies ranging 78.03\% to 89.85\% across the five fungal isolates (Fig.~\ref{fig:multiarray_performance}a). The highest recognition rate was obtained for \textit{B. adusta} (89.85\%), followed by \textit{X.510} (86.08\%), \textit{P.513} (85.19\%), and \textit{T.508} (85.11\%), whereas \textit{P. toxicarium} showed the lowest accuracy (78.03\%). Some of the confusions observed with a single array persisted, including partial overlap between the two \textit{Penicillium} isolates and cross-genus confusion involving \textit{X.510}. As discussed earlier, these errors are consistent with partially overlapping fungal VOC signatures and the cross-sensitive nature of the sensing array. Nevertheless, the consistently higher class-wise accuracies show that simultaneous observations from spatially separated locations disambiguate chemical signatures that are difficult to distinguish locally.

Across five independent training runs, class-wise precision, recall, and F1-score remained broadly consistent (Fig.~\ref{fig:multiarray_performance}c). \textit{P. toxicarium} continued to exhibit relatively high precision but lower recall, whereas \textit{B. adusta} maintained strong recall across runs. Spatial context therefore reduced, but did not eliminate, ambiguity associated with overlapping fungal VOC responses.

The effect was more pronounced for source direction. Directional classification accuracy ranged from 78.29\% to 87.31\% across the four spatial bins (Fig.~\ref{fig:multiarray_performance}b). Bin~3 achieved the highest class-wise accuracy at 87.31\%, followed by Bin~1 at 85.58\% and Bin~2 at 85.13\%, whereas Bin~0 showed the lowest accuracy at 78.29\%. Misclassifications remained concentrated primarily between neighboring or geometrically adjacent directional bins, indicating that directional ambiguity is still shaped by the spatial continuity of the measured chemical field.

Performance across independent training runs showed relatively stable recall but greater variation in precision among directional bins (Fig.~\ref{fig:multiarray_performance}d). These results demonstrate that spatially distributed chemical measurements substantially improve observability of source direction and also strengthen source recognition. More generally, they show that ambiguity in individual chemical observations can be reduced by acquiring the same VOC field from multiple spatial viewpoints.

\subsection{Sequential chemical measurements enable autonomous closed-loop source localization}

The preceding experiments establish that a local Scensory observation provides uncertain estimates of source identity, direction, and distance, while spatial diversity reduces this uncertainty. A moving robot can acquire such evidence sequentially. We next evaluated whether successive single-array predictions could be accumulated during robot motion to support autonomous closed-loop source localization. 

We evaluated the mobile system in four indoor configurations with different fungal-source placements (Fig.~\ref{fig:realworld_mobile}). For each configuration, we show two runs following predefined patrol routes in opposite directions. During patrol, the robot repeatedly acquired 3\,s VOC observations and inferred target identity, relative direction, and distance. Individual predictions were not converted directly into navigation goals. Instead, accepted predictions generated spatially distributed evidence in a common world frame, allowing hypotheses supported from multiple robot positions and headings to reinforce one another. As evidence accumulated, persistent peaks in the map generated candidate source locations. The robot then interrupted its patrol and actively acquired additional measurements from different headings to verify and refine the candidate. Verified candidates triggered source-directed navigation, during which the estimate continued to update. At the final standoff position, the arm-mounted sensor array was extended toward the candidate for local chemical confirmation. The detailed algorithms and procedures can be found in Sec.~\ref{sec:method-sequential-localization}.

\begin{figure}[htbp]
    \centering
    \includegraphics[width=1\linewidth]{figures/realWorldMobile.pdf}
    \caption{$|$ \textbf{Sequential chemical observations support autonomous closed-loop source localization.}
    Eight selected mobile-robot runs are shown across four indoor configurations with different fungal-source locations. Each row shows one experimental setting with a distinct environment configuration and fungal-source placement, with two selected runs under the same setting. Selected world-frame evidence maps are shown above the corresponding robot trajectories, illustrating how sequential local observations accumulate into spatial source hypotheses as the robot moves through the environment. The four target coordinates were $(1.900,0.700)$\,m, $(1.800,-0.300)$\,m, $(0.100,1.900)$\,m, and $(3.700,0.800)$\,m. The left column shows runs using the predefined counterclockwise patrol trajectory, whereas the right column shows runs using the corresponding clockwise trajectory.}
    \label{fig:realworld_mobile}
\end{figure}

Across the eight selected runs, the planar endpoint error between the final end-effector position and the target was $0.606\pm0.294$\,m (mean $\pm$ sample standard deviation), with individual errors ranging from 0.187 to 1.005\,m (Eq.~\ref{eq:ee-source-error}). This metric evaluates the complete perception--action pipeline and therefore incorporates uncertainty from species recognition, directional inference, range estimation, spatial evidence accumulation, and robot navigation.

System behavior also illustrates why sequential evidence accumulation is important for chemical perception. Ambient airflow was neither actively controlled nor modeled, and transient predictions occurred during several runs. This experiment was also conducted in a different location and season from the original data collection. Rather than immediately committing to these detections, the robot paused and rotated to acquire additional observations. Many transient hypotheses weakened during verification, after which the robot resumed patrol, whereas predictions near the true source became progressively more persistent across successive viewpoints. These experiments demonstrate how robot motion can convert uncertain local chemical observations into spatial evidence sufficient to support candidate rejection, source-directed navigation, and local confirmation.

\subsection{Sensor ablations provide insight into the chemical inference process}

\begin{figure}[htbp]
    \centering
    \includegraphics[width=1\linewidth]{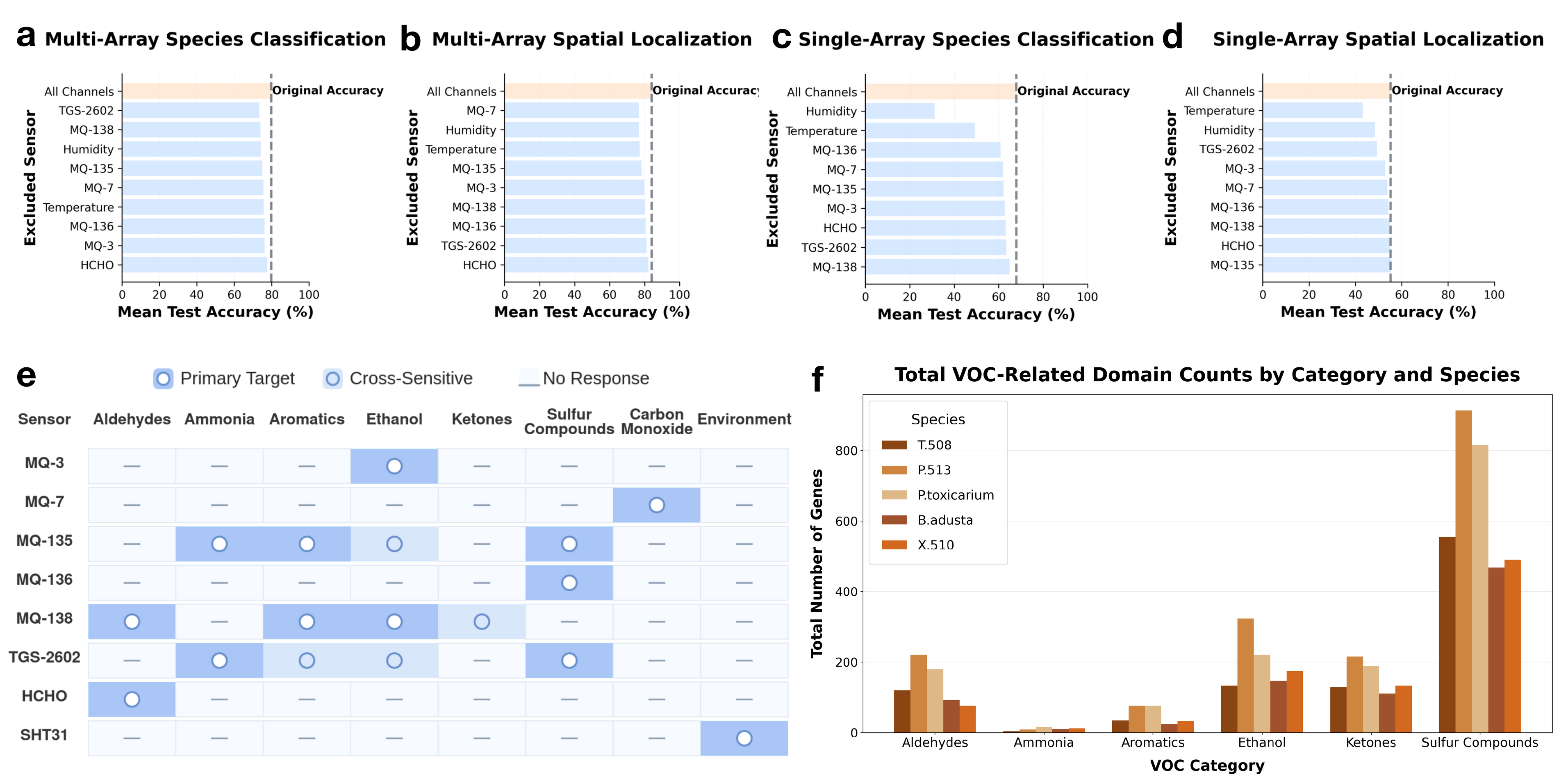}
    \caption{$|$ \textbf{Sensor-channel ablation and biological annotation analysis.}
    \textbf{a-d}, Leave-one-channel-out ablation experiments for spatial-context and single-array models. Performance after removing each input channel is averaged across five independent training runs. Dashed lines indicate performance using all channels. Sensors are ordered according to their contribution to model performance.
    \textbf{a}, Spatial-context species classification.
    \textbf{b}, Spatial-context direction classification.
    \textbf{c}, Single-array species classification.
    \textbf{d}, Single-array direction classification.
    \textbf{e}, Nominal chemical sensitivities of individual sensing channels.
    \textbf{f}, InterPro domain counts grouped into VOC-related functional categories using curated keyword mappings (SI~\ref{supp-note:voc-keywords}). Sulfur-associated annotations were prevalent across the fungal isolates and are shown alongside the importance of sulfur-sensitive sensing channels identified by the ablation analysis.}
    \label{fig:ablation_result}
\end{figure}

We next examined which sensing channels contributed most strongly to the learned semantic and spatial predictions. Leave-one-channel-out ablations were performed for both the single-array and spatial-context models by removing each input channel and measuring the resulting change in performance relative to the full model (Fig.~\ref{fig:ablation_result}a--d).

Environmental measurements had a particularly strong effect in the single-array condition. For species classification, the measured effects associated with temperature and humidity averaged 27.79\% (18.72\% for temperature and 36.85\% for humidity), compared with 6.04\% in the spatial-context condition (5.09\% and 6.98\%, respectively). For directional inference, the corresponding values averaged 9.30\% in the single-array setting (12.01\% for temperature and 6.59\% for humidity) and 7.02\% with spatial context (6.79\% and 7.25\%). The stronger dependence on environmental channels during local inference indicates that temperature and humidity provide useful context when spatial observations are unavailable. This is consistent with the known sensitivity of MOx sensor responses to ambient temperature and relative humidity \cite{sukhdev2022simplified}.

The ablation analysis also identified several sulfur-sensitive channels, including MQ-135, MQ-136, and TGS-2602, as important contributors to fungal species classification (Fig.~\ref{fig:ablation_result}a,c,e). To examine whether this model-derived pattern corresponded to plausible biological differences among the isolates, we compared the sensor importance with InterPro annotations from the fungal isolates. Protein domains were grouped into VOC-related functional categories using curated keyword mappings (SI~\ref{supp-note:voc-keywords}), and sulfur-associated annotations were among the most prevalent categories across isolates (Fig.~\ref{fig:ablation_result}f).

This correspondence does not identify the volatile compounds responsible for individual sensor responses, nor does it establish a direct relationship between specific genomic annotations and measured VOCs. It does, however, provide a biologically plausible context for the prominence of sulfur-sensitive channels in the learned models and motivates targeted chemical characterization in future work. Notably, our ablation studies can reveal such information purely based on computational experiments. Since all models were parallelized on graphics processing units (GPUs), such experiments could be conducted very efficiently.


\section{Discussion}\label{sec3}

Scensory demonstrates that airborne chemical sensing can support a broader form of robotic perception in which a robot infers both the identity and spatial origin of a biological source and uses this information to guide action. From short local VOC observations, the system estimates fungal identity, coarse source direction, and source distance, then integrates successive predictions acquired during motion into a common spatial representation. This enables candidate generation, verification, navigation, and local confirmation within a closed-loop perception--action process. More broadly, the results show that olfactory signals can provide actionable spatial information for autonomous robots.

Recovering spatial information from chemical measurements is challenging because the relationship between a local sensor response and its source geometry is indirect. Cross-sensitive gas sensors measure overlapping responses to mixtures of compounds, while chemical transport and sensor dynamics further shape the temporal signal observed at any location. Scensory learns this relationship through robot-generated spatial supervision. Automated data collection associates temporal VOC responses with source identity, relative direction, and distance. Our single-array experiments show that these local observations contain recoverable semantic and spatial information, while the spatial-context experiments with six distributed arrays show that observations from multiple locations substantially reduce ambiguity. This suggests that the observability of chemical sources depends not only on the sensing hardware but also on how measurements are acquired across space and how the resulting information is represented and learned.

Robot motion provides a complementary way to acquire this spatial information. Without relying on multiple simultaneously deployed arrays, a mobile robot can sample the chemical environment from different positions and orientations over time. In Scensory, local predictions are transformed into a common world frame and accumulated as spatial evidence, allowing consistent observations from different robot poses to reinforce a source hypothesis. Additional measurements acquired during candidate verification can, in turn, suppress transient hypotheses before the robot commits to an approach. The mobile experiments demonstrate this process in a closed loop. Repeated sensing and motion transform individually uncertain chemical observations into persistent spatial evidence that can guide source-directed behavior. This connection between sensing and motion also motivates active olfactory perception, in which future robots could deliberately select positions, orientations, and trajectories that are most informative for resolving competing source hypotheses.

The sensor-level analyses provide additional insight into the information used for these predictions. Temperature and humidity had a stronger influence on species inference in the single-array setting than when measurements from multiple locations were available, indicating that environmental context can become particularly important when inference relies on a local chemical observation. This observation is consistent with the known dependence of MOx sensor responses on ambient temperature and relative humidity \cite{sukhdev2022simplified}. Several channels with nominal sensitivity to sulfur-containing compounds, including MQ-135, MQ-136, and TGS-2602, were also important for fungal species classification. Sulfur-associated functional annotations were prevalent across the fungal isolates, providing plausible biological context for this model-derived pattern. This correspondence does not establish which VOCs produced individual sensor responses or a direct relationship between particular genomic annotations and measured chemical signals. Targeted chemical characterization will be needed to establish these links and could ultimately inform the design of sensing arrays optimized for particular biological sources.

Several aspects of the present study can be improved in future work. Local inference was evaluated on five fungal isolates and on observations with mean source distances between 0.1 and 0.65\,m. Although data collection extended over 125 days, culture age and growth stage were not systematically varied as independent experimental factors. Generalization across growth stages, source strengths, mixed fungal populations, background chemical environments, and different airflow conditions therefore remains to be established. The mobile experiments further comprise eight runs across four indoor configurations, and the reported endpoint errors characterize this subset, which leaves large-scale environmental studies as a valuable future direction. Larger evaluations with randomized source locations, environmental conditions, and previously unseen biological samples will be important for validating the robustness and operating range.

Olfaction ultimately provides robots with access to properties of the environment that are difficult or impossible to observe through geometry and appearance alone. Biological activity, contamination, emissions, and other chemical processes can produce informative signals without corresponding visual, acoustic, or tactile signatures as in current robotic systems. Our results presented here show that these signals can be connected to spatial reasoning and autonomous behavior. Local VOC dynamics provide information about both source identity and relative location, and robot motion provides additional observations that help reduce ambiguity over time. By bringing chemical perception into the perception--action loop, Scensory provides a step toward robots that can autonomously investigate environments not only through their physical structure, but also through their underlying chemical and biological processes.

\section{Methods}\label{sec4}

\subsection{Sensor array and hardware architecture}\label{sec:method-sensor-array}

\begin{figure}
    \centering
    \includegraphics[width=1\linewidth]{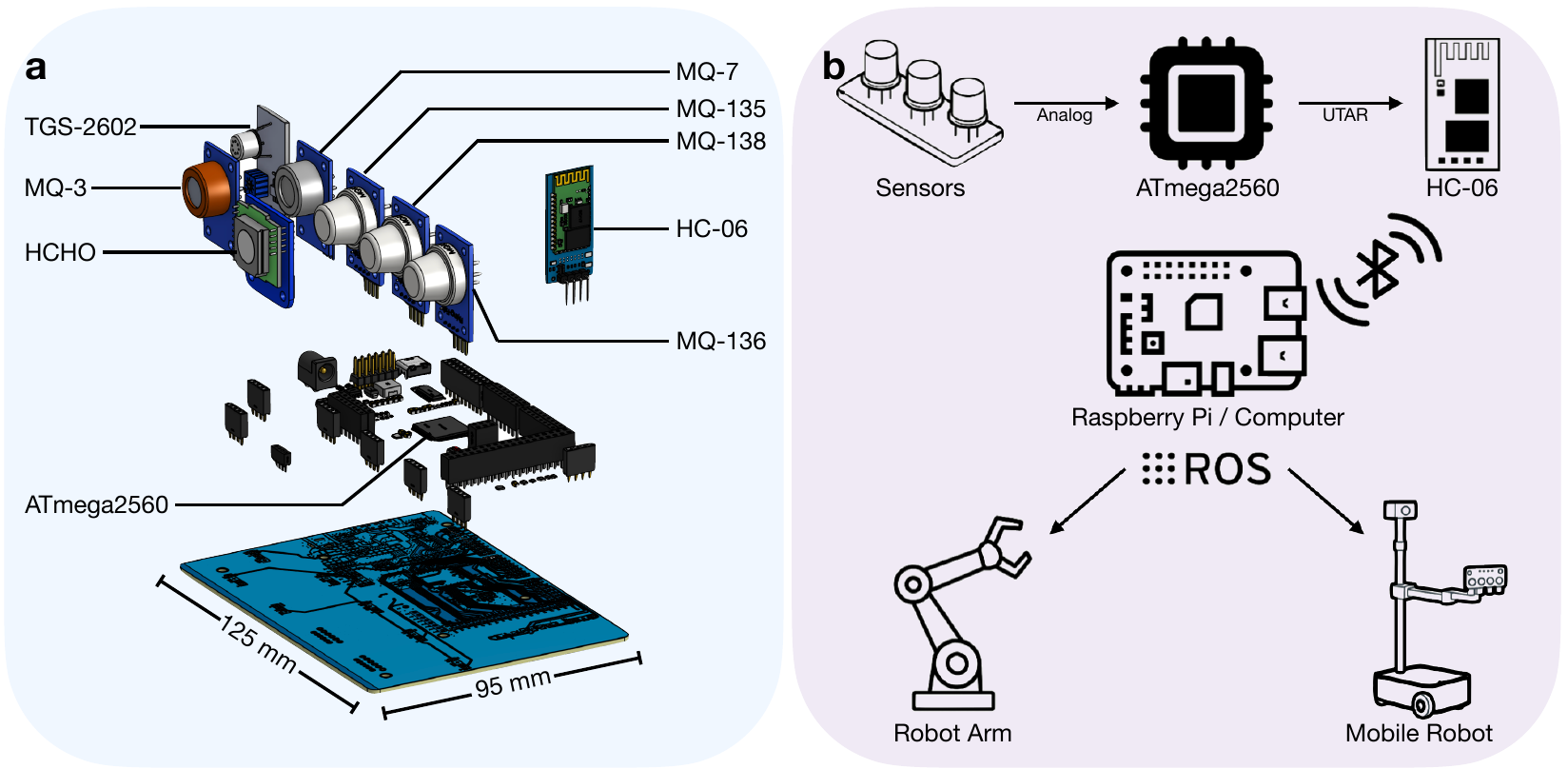}
    \caption{$|$ \textbf{Hardware architecture and wireless communication pipeline of the Scensory array.}
    \textbf{a.} Exploded view of the custom dual-layer PCB sensor array (95 $\times$ 124\,mm), comprising six MOx gas sensors, an electrochemical formaldehyde sensor, temperature and humidity sensors, an ATmega2560 microcontroller, and an HC-06 Bluetooth module for wireless communication.
    \textbf{b.} System-level schematic of data flow: analog sensor signals are digitized by the microcontroller and wirelessly transmitted via Bluetooth to a host computer (Raspberry Pi 5) running ROS, which integrates sensory data for real-time inference and robot control.}
    \label{fig:deviceCommunication}
\end{figure}

The Scensory array integrates six commercially available metal-oxide semiconductor (MOx) gas sensors (MQ-3, MQ-7, MQ-135, MQ-136, MQ-138, and TGS2602), an electrochemical formaldehyde sensor (HCHO), and an SHT31 sensor for ambient temperature and relative humidity (Fig.~\ref{fig:deviceCommunication}a). These sensors have overlapping nominal sensitivities to chemical classes including alcohols, sulfur-containing compounds, carbon monoxide, and aldehydes, providing complementary response patterns to complex VOC mixtures.

All components are surface-mounted on a 95 $\times$ 124\,mm FR-4 PCB with a shared layout, firmware, and calibration protocol. Analog sensor outputs are read by an onboard ATmega2560 microcontroller and transmitted via an HC-06 Bluetooth module to a Raspberry Pi 5 running the Robot Operating System (ROS), which handles communications and robot motion control (Fig.~\ref{fig:deviceCommunication}b). Sensor data are recorded at 10\,Hz and timestamp-synchronized with robot pose information through ROS, pairing each sensor reading with the corresponding source position during data collection.

\subsection{Fungal sample preparation}\label{sec:method-sample-preparation}

All fungal isolates were grown on Potato Dextrose Agar (PDA; IBI Scientific, Dubuque, IA). PDA medium was prepared by dissolving 39 g of powder in 1 L of deionized water and autoclaving at 121$^{\circ}$C and 15 psi for at least 20\,min. Aliquots of 12--15 mL were dispensed into 90\,mm-diameter sterile Petri dishes.

The isolates \textit{B. adusta} and \textit{P. toxicarium} were obtained by swabbing the basement of the PreMiEr Home @ Duke testbed with sterile PBS-moistened swabs. Samples were streaked onto PDA and allowed to grow for five days. Colonies with fungal morphology underwent three rounds of purification by excising regions with similar morphology and transferring them to fresh PDA plates. During purification, \textit{B. adusta} and \textit{P. toxicarium} were incubated at 30$^{\circ}$C for 4 and 7 days, respectively.

Isolates \textit{T.508}, \textit{X.510}, and \textit{P.513} were previously recovered from creosote-contaminated soil samples. Soil and root samples were collected from and around \textit{Spartina alterniflora} and \textit{Phragmites australis} in sterile plastic bags and transported on ice. Roots were rinsed in distilled water and surface-sterilized with 3\% hydrogen peroxide for 1 min. Soil samples were serially diluted to 1:100 and 1:1000 using 1 g of soil from each sampling location. Surface-sterilized roots and diluted soil samples were plated onto Difco Malt Extract Agar (Difco$^{\mathrm{TM}}$, Detroit, MI), Tea Leaf Agar, and Nitrogen Free Agar. Plates were examined after 10 days of growth. Phenotypically distinct fungal colonies were transferred onto fresh malt extract agar and subjected to three rounds of plate purification.

For identification, the ITS region was amplified and the amplicons were submitted for Sanger sequencing. Sequences were compared with the NCBI database using BLASTn. The reported sequence similarities were 98.26\% for \textit{B. adusta}, 98.05\% for \textit{P. toxicarium}, 98.12\% for \textit{T.508}, 98.07\% for \textit{X.510}, and 86.59\% for \textit{P.513}.

\subsection{Robot-enabled spatial supervision and automated data collection}\label{sec:method-platform}

Training and offline evaluation data were collected in a custom-built acrylic testbed measuring 1.5\,m $\times$ 1.0\,m $\times$ 0.6\,m (L $\times$ W $\times$ H), with a total volume of 0.9\,m\textsuperscript{3}. The testbed remained sealed during measurement intervals to reduce uncontrolled external airflow and improve experimental repeatability. A side-mounted exhaust fan remained off during measurements and was activated between trials to purge residual VOCs. Data collection resumed only after sensor readings returned to baseline.

Each cultured sample was attached to the end effector of a WidowX 200 robotic arm, which moved the source through a predefined sequence of spatial waypoints. At each waypoint, the arm remained stationary for 10--20\,s while six fixed sensor arrays recorded synchronously. The robot end-effector pose provided source-location ground truth for the synchronized sensor recordings. For single-array learning, each array's measurements were paired with source labels expressed in its local coordinate frame. The same recordings also supported multi-array analysis by combining simultaneous measurements from all six sensing locations.

\subsection{Data preprocessing and label construction}\label{sec:method-data-preprocessing}

Sensor measurements and robot trajectories were continuously recorded in ROS bag files and converted into timestamp-aligned CSV files. Timestamps were floored to 100-ms intervals to align sensor measurements with their nearest end-effector pose. A common preprocessing pipeline was then applied to the recorded sensor streams. MOx channels were first smoothed using a moving-average filter with a window size of 10, followed by Savitzky--Golay filtering \cite{savitzky1964smoothing} with a window size of 41 and polynomial order of 2. Baseline correction was performed by subtracting the mean of the first five filtered samples, followed by a uniform positive shift to ensure non-negative values. Temperature, humidity, robot position, and source distance were preserved unchanged during this filtering and baseline-correction stage.

The processed time series were segmented into 3-s windows for fungal-species classification, source-direction classification, and distance regression. Species labels were assigned according to the fungal isolate used in each experiment.

For the single-array models, each array was processed independently. For an array located at $(x_s,y_s,z_s)$ with orientation $\theta_s$, the source position was transformed from the global frame into the array-local frame as
\begin{equation}
\begin{bmatrix}
x' \\ y' \\ z'
\end{bmatrix}
=
R(\theta_s)
\left(
\begin{bmatrix}
x \\ y \\ z
\end{bmatrix}
-
\begin{bmatrix}
x_s \\ y_s \\ z_s
\end{bmatrix}
\right),
\end{equation}
where $R(\theta_s)$ denotes the rotation matrix from the global frame to the array-local frame and $[x,y,z]^T$ is the source position in the global frame. Direction labels were assigned by partitioning the local $x$--$y$ plane into four quadrants. Source distance was computed as the Euclidean distance between the sensor-array center and the robot end effector, with the final time step of each window used as the regression target.

For each single-array stream, windows with source bearings within 10$^{\circ}$ of either local coordinate axis were excluded to reduce ambiguity near directional class boundaries. For the single-array models, only windows with stable robot poses, defined as less than 5\,cm of positional variation over at least 7\,s, were retained. Local-model windows were further retained only when their mean source-to-array distance was greater than 0.1\,m and less than 0.65\,m. The instantaneous regression target could fall outside this window-mean selection interval.

For the multi-array models, synchronized readings from all six arrays were concatenated into a joint input, and directional labels were defined in the global testbed frame. For these multi-array models, windows with source bearings within 10$^{\circ}$ of either global coordinate axis were excluded, applying the same boundary-filtering criterion in the global frame.

Training and validation samples were constructed according to the task-specific procedures described in Supplementary Note~\ref{supp-note:model-architectures}. The test sets were derived from separate experimental trials collected at different times or on different days and were not used for model training or hyperparameter selection. Parameters for all preprocessing steps were estimated on the training data and subsequently applied to the validation and test sets. Multi-array input features were standardized using StandardScaler, whereas the single-array distance pipeline used MinMaxScaler with a nominal feature range of $[0,1]$. Sampling and loss weighting for class imbalance were applied during training, as detailed in the task-specific training settings.

\subsection{Task-specific models for local and spatial-context inference}\label{sec:method-neural-architecture}

Scensory uses task-specific neural networks to estimate species identity, source direction, and source distance from short VOC time series, with each architecture matched to its prediction task. Joint identification and localization combines these model outputs at the system level. All models operate on the same nine-channel representation per sensor array, and no airflow, vision, or other exteroceptive measurements are provided as inputs.

For the primary single-array local setting, three model families estimate fungal species, relative source direction, and source distance from local VOC observations. The species classifier uses two residual temporal-convolution blocks \cite{bai2018empirical} to encode short-term sensor dynamics, followed by sinusoidal positional encoding and four-head temporal self-attention \cite{vaswani2017attention}, global temporal pooling, and an MLP classification head. This architecture is intended to capture discriminative temporal response patterns across the cross-sensitive sensing channels.

Relative source direction is estimated using a separate four-class classifier. The direction model uses a three-layer bidirectional long short-term memory (BiLSTM) encoder to model temporal dependencies in the local VOC sequence, followed by positional encoding, four-head self-attention, mean pooling, and an MLP output head. The four output classes correspond to coarse directional bins in the local sensor frame, as defined in Sec.~\ref{sec:method-data-preprocessing}.

Source distance is estimated using species-specific regression models. A separate distance regressor is trained for each fungal species, yielding five independently parameterized regressors. Each regressor uses a two-layer BiLSTM followed by two self-attention stages, learned temporal-attention pooling, residual feature projection, and a scalar regression head. At deployment, species prediction identifies the source class, and distance estimation uses the regressor for the species being localized. This decomposition allows each regressor to model species-specific differences in the relationship between temporal VOC responses and source proximity.

In addition to the local models, we train two spatial-context reference models using synchronized measurements from all six sensor arrays. These models are used only to evaluate how simultaneous spatial redundancy affects chemical observability and are not part of the mobile deployment architecture. At each time step, the spatial-context species classifier applies a shared two-block convolutional encoder across the ordered six-array dimension, followed by two self-attention stages over that same dimension. The representations are then averaged over time. Sinusoidal positional encoding over array positions and mean pooling across arrays produce the feature vector used by a two-layer MLP species head. The spatial-context direction classifier instead preserves array identity through six separate convolutional branches. Each branch contains two convolution--normalization--pooling stages and produces a 32-dimensional embedding; the six embeddings are concatenated and passed to an MLP that predicts one of four global direction bins.

Detailed layer-wise architectures and training settings for all models are provided in Supplementary Note~\ref{supp-note:model-architectures} (Tables~\ref{supp-table:multi-species-architecture}--\ref{supp-table:single-distance-architecture}).

\subsection{Closed-loop fungal source localization}\label{sec:method-sequential-localization}

The real-world system was implemented in ROS on a Hello Robot Stretch. Before each experiment, a 2D occupancy map and a predefined patrol route were created. During deployment, the robot followed the patrol route while the navigation stack performed online obstacle avoidance. During patrol, the onboard Scensory array published $\mathbf{z}_t=[a_{0,t},\ldots,a_{6,t},h_t,T_t]^\top$. These streaming measurements were processed using a pipeline adapted from the single-array preprocessing framework to reduce short-term noise and gas-sensor baseline drift before model inference. The seven gas-sensor channels were combined with the uncorrected humidity and temperature measurements to form the nine-channel model input. A short startup interval was excluded before accepting predictions so that initialization and baseline transients did not contribute to the persistent map.

At each subsequent sensor update, the latest $W=30$ samples formed $\mathbf{X}_t\in\mathbb{R}^{W\times9}$. The local models returned a species distribution $\mathbf{q}_t\in[0,1]^5$, a direction distribution $\boldsymbol{\pi}_t\in[0,1]^4$, and a range estimate $\hat d_t$. In the indoor mobile-robot experiments, $c^*$ denoted the target species, \textit{B. adusta}, and the controller used the corresponding species-specific distance regressor. The four direction-bin centers in the robot base frame were $\boldsymbol{\alpha}=[135^\circ,45^\circ,-135^\circ,-45^\circ]$, where $0^\circ$ denotes forward and positive angles rotate counterclockwise.

We accumulated accepted predictions on a 12 $\times$ 12\,m world-frame grid with 0.05-m resolution. Let the robot pose at update $t$ be $\mathbf{s}_t=(x_t,y_t,\psi_t)$ and let a grid-cell center be $\mathbf{g}=(x,y)$. Its range and robot-relative bearing are
\begin{align}
r_t(\mathbf{g}) &= \sqrt{(x-x_t)^2+(y-y_t)^2},\\
\phi_t(\mathbf{g}) &= \operatorname{wrap}\!\left[\operatorname{atan2}(y-y_t,x-x_t)-\psi_t\right].
\end{align}
One prediction contributes the spatial evidence field
\begin{equation}
L_t(\mathbf{g})=
\exp\!\left[-\frac{(r_t(\mathbf{g})-\hat d_t)^2}{2\sigma_r^2}\right]
\sum_{k=1}^{4}\pi_{t,k}
\exp\!\left[-\frac{\operatorname{wrap}(\phi_t(\mathbf{g})-\alpha_k)^2}{2\sigma_\phi^2}\right],
\label{eq:spatial-evidence}
\end{equation}
with $\sigma_r=0.10$\,m and $\sigma_\phi=25^\circ$. Thus, each observation produces an annular angular distribution rather than a single projected endpoint. The complete four-bin probability vector is retained, allowing evidence acquired at different robot poses to reinforce intersecting hypotheses.

An individual prediction was considered valid when the target-species probability and maximum direction-bin probability met the pre-specified thresholds $\tau_s=0.485$ and $\tau_d=0.425$, respectively:
\begin{equation}
q_{t,c^*}\geq\tau_s, \qquad \max_k\pi_{t,k}\geq\tau_d.
\end{equation}
The confidence thresholds were specified before deployment and were not adjusted online. Accepted observations updated an unnormalized evidence map according to
\begin{align}
H_t(\mathbf{g}) &= \lambda H_{t-1}(\mathbf{g})+
w_t\frac{L_t(\mathbf{g})}{\sum_{\mathbf{g}'\in\mathcal{G}}L_t(\mathbf{g}')},
\label{eq:evidence-update}\\
w_t &= q_{t,c^*}\max_k\pi_{t,k},
\end{align}
where $\mathcal{G}$ is the map grid and $\lambda=0.985$ is applied at each accepted update. This map is an additive confidence representation, not a calibrated Bayesian posterior.

The candidate source was the grid maximum $\hat{\mathbf{g}}_t=\arg\max_{\mathbf{g}}H_t(\mathbf{g})$. We quantified peak concentration by
\begin{equation}
C_t(\rho)=
\frac{\sum_{\|\mathbf{g}-\hat{\mathbf{g}}_t\|_2\leq\rho}H_t(\mathbf{g})}
{\sum_{\mathbf{g}\in\mathcal{G}}H_t(\mathbf{g})}.
\label{eq:peak-concentration}
\end{equation}
A source candidate became eligible for verification only after repeated accepted observations produced a sufficiently concentrated peak, so a single prediction could not directly initiate source approach. The robot then paused its patrol, collected a stationary verification window, and performed bounded in-place rotations to acquire additional observations from different sensor headings. The consistency of the target-species, direction, and projected-position estimates was used to refine the source hypothesis. Verified candidates were converted into collision-aware navigation goals positioned at a safe standoff from the estimated source, and the source estimate continued to update during approach. At the standoff pose, the robot refined its sensing orientation, extended the arm-mounted array, and performed local chemical confirmation before terminating the run. Safety checks and loss-of-evidence conditions returned the robot to its saved patrol when verification, navigation, or local confirmation was unsuccessful. Controller recovery and local-confirmation details are provided in Supplementary Note~\ref{supp-note:closed-loop-controller}.

For each real-world run, we define the 2D planar endpoint error as
\begin{equation}
e_{\mathrm{EE},\mathrm{2D}}^{(i)}=
\sqrt{
\left(x_{\mathrm{EE},i}^{\mathrm{final}}-x_{\mathrm{target},i}\right)^2+
\left(y_{\mathrm{EE},i}^{\mathrm{final}}-y_{\mathrm{target},i}\right)^2
},
\label{eq:ee-source-error}
\end{equation}
where $(x_{\mathrm{EE},i}^{\mathrm{final}},y_{\mathrm{EE},i}^{\mathrm{final}})$ is the final end-effector position and $(x_{\mathrm{target},i},y_{\mathrm{target},i})$ is the measured target position for run $i$, both expressed in the common map frame. This metric reports the 2D planar endpoint error of the complete perception-and-navigation system and does not include vertical displacement.

\subsection{Biological annotation analysis}\label{sec:method-interpor-scan}

To examine whether model-derived sensor importance was consistent with known fungal metabolic functions, we performed genomic sequencing and functional annotation of the five isolates using InterPro scan. To prepare samples for genomic analysis, a 1 cm by 1 cm cutting of each previously prepared isolate plate was transferred to a new PDA plate and incubated for 4 to 7 days, depending on the species. Once the fungal plates were grown, 250 ng of the fungal growth on the agar was cut out, and the DNA was extracted using a DNeasy Soil Power Pro kit, sent for Sanger sequencing using ITS/LSU primers \cite{sting2019rapid}, and prepared for long-read sequencing using the Rapid Barcoding Kit 24 V14 (SQK-RBK114.24). The isolates were sequenced on a PromethION 2 Solo using a Flow Cell R10 (M Version) (FLO-PROM114M) and monitored using MinKNOW24.11.10. Dorado (v1.0.2) was used to basecall Nanopore pod5 files into fastq.gz files \cite{dorado2025}. These files were assembled into contigs with Flye (v2.9.5) \cite{kolmogorov2019assembly} and polished with Medaka(v2.0.1) before using the Funannotate (v1.8.16) annotation software. All isolate assemblies were processed with this software to obtain gene predictions, annotations, and comparisons between isolates. InterProScan (v5.73-104.0) was used to identify protein domains across species \cite{jones2014interproscan,blum2025interpro}. Domains assigned to VOC-related functional categories were counted and compared descriptively with nominal sensor sensitivities (Fig.~\ref{fig:ablation_result}e,f).


\section{Acknowledgments}\label{sec5}

\noindent\textbf{Funding:} This work is supported by NSF Engineering Research Center for Precision Microbiome Engineering (PreMiEr) under award 2133504. \\

\noindent\textbf{Author contributions:} B.C. and Y.L. conceived and designed the research. Y.L. and E.B. designed and performed physical experiments. All authors analyzed data and wrote the manuscript. \\

\noindent\textbf{Competing interests:} Duke University has filed patent rights for the technology associated with this manuscript. For further license rights, including using the patent rights for commercial purposes, please contact Duke's Office for Translation and Commercialization (otcquestions@duke.edu) and reference OTC DU8859PROV. \\

\noindent\textbf{Data and code availability}: All datasets, open-source code, and trained models are available at \href{https://github.com/generalroboticslab/Scensory}{https://github.com/generalroboticslab/Scensory}. \\




\bibliography{sn-bibliography}


\newpage
\section*{Supplementary Information}

\newcounter{suppitem}
\renewcommand{\thesuppitem}{\Alph{suppitem}}

\refstepcounter{suppitem}
\subsection*{\thesuppitem. VOC category keyword mapping for InterProScan annotation}
\label{supp-note:voc-keywords}

To annotate InterProScan protein domain results with putative volatile organic compound (VOC) functions, we defined the following keyword lists for each VOC category. Any InterProScan domain whose description contained one or more of these keywords was assigned to the corresponding VOC group during domain counting.\\

\begin{itemize}
    \item \textbf{Aldehydes}: aldehyde, aldehyde dehydrogenase, aldehyde oxidase, formaldehyde, formaldehyde dehydrogenase, formaldehyde-activating, formaldehyde lyase, formaldehyde oxidase, formaldehyde transketolase, methylotrophy, S-(hydroxymethyl)glutathione, HMGSH
    \item \textbf{Ammonia}: ammonia, ammonium, ammonia monooxygenase, ammonia transporter
    \item \textbf{Aromatics}: aromatic, phenyl, benzene, phenol, toluene, xylene, catechol, naphthalene
    \item \textbf{Ethanol}: ethanol, alcohol dehydrogenase
    \item \textbf{Ketones}: ketone, acetyl, acetone, acetoacetate
    \item \textbf{Sulfur Compounds}: sulfur, cysteine, methionine, methyltransferase, thio, thiol, sulfide, sulfonate, sulfatase
\end{itemize}

\refstepcounter{suppitem}
\subsection*{\thesuppitem. Model architectures and hyperparameters}
\label{supp-note:model-architectures}

All models receive nine channels per array: seven gas-sensor channels, humidity, and temperature. At the 10-Hz sampling rate, each 3-s input window contains 30 time steps. Tensor dimensions below omit the batch dimension. All models were trained for at most 500 epochs with early-stopping patience of 20 epochs and were evaluated using the same five random seeds: 860, 6265, 15795, 54886, and 76820. Multi-array species and direction models used approximately 80/20 and 70/30 training/validation splits, respectively, with samples randomly reassigned for each seed. Single-array models used fixed, approximately 70/30 splits across seeds. Single-array species and direction models used 8,380 training and 3,592 validation windows; distance models used a separate pool of 6,440 training and 2,759 validation windows, filtered by species for each regressor.

\Needspace{0.48\textheight}
\subsubsection*{Multi-array species classification model}

\begin{table}[htbp]
\centering
\caption{\textbf{Multi-array species classification architecture.} At each time step, convolution and two self-attention stages operate across the ordered six-array dimension before temporal averaging. Dimensions follow execution order.}
\label{supp-table:multi-species-architecture}
\begin{footnotesize}
\begin{tabular}{@{}p{0.16\linewidth}p{0.44\linewidth}p{0.14\linewidth}p{0.14\linewidth}@{}}
\toprule
\textbf{Module} & \textbf{Operation} & \textbf{Parameters} & \textbf{Output size} \\
\midrule
Input & Synchronized data organized as time $\times$ arrays $\times$ channels & 9 channels per array & $30\times6\times9$ \\
Array encoder block 1 & Two Conv1D--BatchNorm--ReLU--dropout layers across arrays, with residual projection & $9\rightarrow64$, $k=5$, $d=1$, $p=4$, dropout 0.3 & $30\times6\times64$ \\
Array encoder block 2 & Two Conv1D--BatchNorm--ReLU--dropout layers across arrays, with residual addition & $64\rightarrow64$, $k=5$, $d=2$, $p=8$, dropout 0.3 & $30\times6\times64$ \\
Array self-attention 1 & Self-attention over six arrays at each time step, residual addition, dropout, and LayerNorm & 8 heads, dropout 0.1 & $30\times6\times64$ \\
Axis transpose & Exchange the time and array axes & -- & $6\times30\times64$ \\
Array self-attention 2 & Self-attention over six arrays at each time step, residual addition, dropout, and LayerNorm & 8 heads, dropout 0.1 & $6\times30\times64$ \\
Temporal aggregation & Mean pooling over time & -- & $6\times64$ \\
Array encoding and pooling & Sinusoidal positional encoding over array positions and mean pooling & -- & $64$ \\
Species head & Linear--ReLU--dropout--linear & $64\rightarrow64\rightarrow5$, dropout 0.3 & $5$ \\
\bottomrule
\end{tabular}
\end{footnotesize}
\end{table}
\FloatBarrier

This model uses cross-entropy loss and the Adam optimizer with an initial learning rate of $1\times10^{-4}$ and batch size 32. No learning-rate scheduler is applied.

\Needspace{0.38\textheight}
\subsubsection*{Multi-array direction classification model}

\begin{table}[htbp]
\centering
\caption{\textbf{Multi-array direction classification architecture.} Each array is processed by an independently parameterized convolutional branch before feature fusion.}
\label{supp-table:multi-direction-architecture}
\begin{footnotesize}
\begin{tabular}{@{}p{0.16\linewidth}p{0.44\linewidth}p{0.14\linewidth}p{0.14\linewidth}@{}}
\toprule
\textbf{Module} & \textbf{Operation} & \textbf{Parameters} & \textbf{Output size} \\
\midrule
Input & Six synchronized preprocessed sensor windows & 9 channels per array & $6\times30\times9$ \\
Branch convolution 1 & Conv1D--BatchNorm--ReLU & $9\rightarrow16$, $k=3$, $s=1$, $p=1$ & $6\times16\times30$ \\
Branch pooling 1 & One-dimensional max pooling & $k=2$ & $6\times16\times15$ \\
Branch convolution 2 & Conv1D--BatchNorm--ReLU & $16\rightarrow32$, $k=3$, $s=1$, $p=1$ & $6\times32\times15$ \\
Branch pooling 2 & One-dimensional max pooling & $k=2$ & $6\times32\times7$ \\
Branch projection & Flatten and independent linear projection for each array & $224\rightarrow32$ & $6\times32$ \\
Array fusion & Concatenation of the six array embeddings & -- & $192$ \\
Direction head & Linear--ReLU--linear & $192\rightarrow64\rightarrow4$ & $4$ \\
\bottomrule
\end{tabular}
\end{footnotesize}
\end{table}
\FloatBarrier

This model uses cross-entropy loss and the Adam optimizer with an initial learning rate of $1\times10^{-5}$ and batch size 64. No dropout or learning-rate scheduler is applied.

\Needspace{0.45\textheight}
\subsubsection*{Single-array species classification model}

\begin{table}[htbp]
\centering
\caption{\textbf{Single-array species classification architecture.} Two dilated residual temporal-convolution blocks encode one local sensor window before attention-based temporal aggregation.}
\label{supp-table:single-species-architecture}
\begin{footnotesize}
\begin{tabular}{@{}p{0.16\linewidth}p{0.44\linewidth}p{0.14\linewidth}p{0.14\linewidth}@{}}
\toprule
\textbf{Module} & \textbf{Operation} & \textbf{Parameters} & \textbf{Output size} \\
\midrule
Input & One preprocessed sensor window & 9 channels & $30\times9$ \\
TCN block 1 & Two Conv1D--GroupNorm--ReLU--dropout layers with residual projection & $9\rightarrow64$, $k=5$, $d=1$, $p=4$, dropout 0.3 & $30\times64$ \\
TCN block 2 & Two Conv1D--GroupNorm--ReLU--dropout layers with residual addition & $64\rightarrow64$, $k=5$, $d=2$, $p=8$, dropout 0.3 & $30\times64$ \\
Block fusion & Element-wise addition of the two TCN block outputs & -- & $30\times64$ \\
Temporal encoding & Sinusoidal positional encoding & dropout 0.1 & $30\times64$ \\
Self-attention & Temporal multi-head self-attention & 4 heads & $30\times64$ \\
Temporal pooling & Mean pooling over time & -- & $64$ \\
Species head & Linear--ReLU--dropout--linear & $64\rightarrow64\rightarrow5$, dropout 0.3 & $5$ \\
\bottomrule
\end{tabular}
\end{footnotesize}
\end{table}
\FloatBarrier

This model uses class-weighted cross-entropy loss and AdamW with weight decay $1\times10^{-4}$, initial learning rate $1\times10^{-4}$, and batch size 32. The learning rate is reduced by a factor of 0.5 after three epochs without improvement in validation accuracy.

\Needspace{0.35\textheight}
\subsubsection*{Single-array direction classification model}

\begin{table}[htbp]
\centering
\caption{\textbf{Single-array direction classification architecture.} Bidirectional recurrent encoding and self-attention map one local sensor window to four array-frame direction bins.}
\label{supp-table:single-direction-architecture}
\begin{footnotesize}
\begin{tabular}{@{}p{0.16\linewidth}p{0.44\linewidth}p{0.14\linewidth}p{0.14\linewidth}@{}}
\toprule
\textbf{Module} & \textbf{Operation} & \textbf{Parameters} & \textbf{Output size} \\
\midrule
Input & One preprocessed sensor window & 9 channels & $30\times9$ \\
Temporal encoder & Three-layer bidirectional LSTM & 64 units per direction, dropout 0.4 & $30\times128$ \\
Temporal encoding & Sinusoidal positional encoding and dropout & dropout 0.4 & $30\times128$ \\
Self-attention & Temporal multi-head self-attention & 4 heads, dropout 0.4 & $30\times128$ \\
Temporal pooling & Mean pooling over time & -- & $128$ \\
Direction head & Linear--GroupNorm--ReLU--dropout--linear & $128\rightarrow64\rightarrow4$, dropout 0.4 & $4$ \\
\bottomrule
\end{tabular}
\end{footnotesize}
\end{table}
\FloatBarrier

This model uses AdamW with weight decay $1\times10^{-4}$, initial learning rate $1\times10^{-4}$, and batch size 64. Training samples are drawn with replacement using weights proportional to inverse bin frequency multiplied by difficulty factors $[1.0,2.0,1.5,1.5]$ for bins 0--3. The loss is the mean per-sample cross-entropy weighted by $[1.0,1.2,1.1,1.0]$, with an additional penalty of 0.1 times the mean Euclidean norm of the logits for Bin~1 samples when present. Before gradient clipping at norm 1.0, gradients are scaled by the batch mean of bin factors $[1.0,1.1,1.0,1.0]$. The learning rate is reduced by a factor of 0.5 after three epochs without improvement in validation accuracy.

\Needspace{0.48\textheight}
\subsubsection*{Species-conditioned single-array distance regression model}

\begin{table}[htbp]
\centering
\caption{\textbf{Species-conditioned single-array distance regression architecture.} Five regressors use the same architecture but are trained independently, one for each fungal species.}
\label{supp-table:single-distance-architecture}
\begin{footnotesize}
\begin{tabular}{@{}p{0.16\linewidth}p{0.44\linewidth}p{0.14\linewidth}p{0.14\linewidth}@{}}
\toprule
\textbf{Module} & \textbf{Operation} & \textbf{Parameters} & \textbf{Output size} \\
\midrule
Input & One preprocessed sensor window & 9 channels & $30\times9$ \\
Input projection & Linear--ReLU--dropout--linear & $9\rightarrow18\rightarrow9$, dropout 0.1 & $30\times9$ \\
Temporal encoder & Two-layer bidirectional LSTM & 128 units per direction, dropout 0.2 & $30\times256$ \\
Temporal encoding & Sinusoidal positional encoding and dropout & dropout 0.2 & $30\times256$ \\
Self-attention 1 & Temporal multi-head self-attention & 8 heads, dropout 0.2 & $30\times256$ \\
Self-attention 2 & Temporal multi-head self-attention with residual addition & 4 heads, dropout 0.2 & $30\times256$ \\
Attention pooling & Learned temporal weights from a two-layer scoring network and weighted sum & $256\rightarrow128\rightarrow1$ & $256$ \\
Shared projection & Linear--GroupNorm--ReLU--dropout--linear--GroupNorm--ReLU--dropout & $256\rightarrow256\rightarrow128$, dropout 0.2 & $128$ \\
Residual projection & Linear projection of the pooled feature and residual addition & $256\rightarrow128$ & $128$ \\
Distance head & Linear--ReLU--dropout--linear--ReLU--dropout--linear & $128\rightarrow128\rightarrow64\rightarrow1$, dropout 0.2 & $1$ \\
\bottomrule
\end{tabular}
\end{footnotesize}
\end{table}
\FloatBarrier

Each species-specific regressor uses mean squared error loss and AdamW with weight decay $1\times10^{-4}$, initial learning rate $1\times10^{-3}$, and batch size 64. The learning rate is reduced by a factor of 0.5 after five epochs without improvement in validation distance error.

\clearpage

\refstepcounter{suppitem}
\subsection*{\thesuppitem. Closed-loop controller recovery and local confirmation}
\label{supp-note:closed-loop-controller}

Candidates detected during system initialization or far from the current robot pose were retained and reconsidered as the patrol acquired additional nearby observations. When a stable candidate became eligible, the navigation controller canceled the current waypoint but retained the patrol index. If stationary verification or active rotational sensing failed, the candidate was temporarily suppressed and the interrupted patrol resumed.

During source approach, the estimate continued to update as new observations became available. Loss of chemical evidence, stale sensor or odometry data, navigation timeout, or excessive approach travel canceled the source goal and returned the robot to the saved patrol. Emergency-stop and runstop events latched the system in a stopped state that required restart.

At the standoff pose, a finer rotational search selected the final sensing orientation. The controller then raised and extended the arm-mounted array and allowed a short settling interval before local confirmation. A valid near-source observation counted as a hit when the target-species and direction predictions were sufficiently confident and the estimated distance was within the local confirmation range. If $M$ valid predictions were collected and $m$ were hits, confirmation required
\begin{equation}
M\geq4, \qquad \frac{m}{M}\geq0.60.
\label{eq:local-confirmation}
\end{equation}
On success, the robot held position. On failure, it retracted the sensor, placed the candidate in cooldown, and resumed the saved patrol route.



\end{document}